\documentclass[letterpaper,twocolumn,english,american,prl, showpacs]{revtex4}
\usepackage[T1]{fontenc}
\usepackage[latin1]{inputenc}
\usepackage{amsmath}
\usepackage{babel}
\usepackage{graphics}
\usepackage{amssymb}

\makeatletter

\providecommand{\LyX}{L\kern-.1667em\lower.25em\hbox{Y}\kern-.125emX\@}
\newcommand{\noun}[1]{\textsc{#1}}

\DeclareMathOperator{\Pre}{Pre}
\DeclareMathOperator{\Post}{Post}
\DeclareMathOperator{\rank}{rank}

\newcommand{\eps}{\varepsilon}
\newcommand{\tauh}{\frac{\tau}{2}}

\newcommand{\hD}[2]{h(#1+D_{i,#2} , \eps_{i j_{#2}})=:
  \beta_{i,#2}}

\makeatother
\begin{document}

\title{Coexistence of regular and irregular dynamics in complex networks
of pulse-coupled oscillators}

\author{Marc Timme, Fred Wolf, and Theo Geisel}

\address{Max-Planck-Institut für Strömungsforschung, 37073 Göttingen, Germany}

\begin{abstract}
For general networks of pulse-coupled oscillators, including regular,
random, and more complex networks, we develop an exact stability analysis
of synchronous states. As opposed to conventional stability analysis,
here stability is determined by a multitude of linear operators. We
treat this multi-operator problem exactly and show that for inhibitory
interactions the synchronous state is stable, independent of the parameters
and the network connectivity. In randomly connected networks with
strong interactions this synchronous state, displaying \textit{regular}
dynamics, coexists with a balanced state that exhibits \textit{irregular}
dynamics such that external signals may switch the network between
qualitatively distinct states. 
\end{abstract}


\maketitle
Complex networks appear as a variety of natural and artificial systems,
ranging from the world wide web and electrical power grids to metabolic
and neural networks \cite{Albert_et_al,Strogatz}. While recent studies
have focussed on their \textit{structure} \cite{Albert_et_al}, the
\textit{dynamics} in such networks constitute a challenging issue
of current and future research \cite{Strogatz}. Even if the individual
vertices of the network are simple dynamical systems, such as limit
cycle oscillators, an exact mathematical analysis of their collective
dynamics is often highly intricate, due to the complex connectivity
structure.

As a prototypical class of dynamical systems interacting on networks,
pulse-coupled units have received a significant amount of interest
because of their relevance to diverse natural systems \cite{Abbott_et_al,Mirollo_et_al,Brunel,Gerstner_et_al,Dror}
including cardiac pacemaker cells, flashing fireflies, earthquakes,
and biological neural networks. Particularly in neuroscience, these
models \cite{Maass_et_al} are essential for understanding collective
dynamic phenomena such as synchronization or the propagation of sensory
signals through extended networks \cite{syn,irreg}. Although biological
neural networks, like other networks occurring in the real world,
often possess a complex connectivity structure, most theoretical studies
on pulse-coupled oscillators are either restricted to networks of
globally coupled oscillators and simple regular networks, or work
in some mean field limit \cite{Abbott_et_al,Mirollo_et_al,Brunel,Dror}.

In this Letter, we study pulse-coupled oscillators interacting on
networks with \textit{general connectivities}, including fully connected,
regular, random, and more complex topologies. We develop an exact
stability analysis of the synchronous state. In contrast to conventional
stability problems, the first order stability operator here is not
linear, but can be expressed by a multitude of distinct linear operators,
the domains of which depend on the rank order of a specific perturbation.
For generally structured networks, the number of operators increases
exponentially with the size of the network. Our analysis provides
a method to treat this multi-operator problem analytically. For networks
with inhibitory couplings, we prove that the synchronous state is
stable, independent of the parameters and the connectivity structure. 

Proceeding from this result, we show that the synchronous state, that
displays regular dynamics, coexists with a state of highly irregular
dynamics in randomly connected networks. We suggest a simple mechanism
for switching between these states. These results establish that the
behavior of networks of pulse-coupled units for a given set of parameters
may be dominated by qualitatively distinct dynamical states.

We consider a system of \( N \) coupled oscillators \cite{Mirollo_et_al}
which interact on directed graphs by sending and receiving pulses.
The structure of this graph is specified by the sets \( \Pre (i) \)
of presynaptic oscillators that send pulses to oscillator \( i \).
These sets determine the sets \( \Post (i) \) of postsynaptic oscillators
that receive pulses from \( i \). A phase-like variable \( \phi _{i}(t)\in (-\infty ,1] \)
specifies the state of each oscillator at time \( t \). The dynamics
of a non-interacting oscillator \( i \) is given by \begin{equation}
\label{eq:phi_dot_1}
d\phi _{i}(t)/dt=1.
\end{equation}
When oscillator \( i \) reaches a threshold, \( \phi _{i}(t)=1 \),
its phase is reset to zero, \( \phi _{i}(t^{+})=0 \), and the oscillator
is said to {}``fire''. A pulse is sent to all postsynaptic oscillators
\( j\in \Post (i) \) which receive this signal after a delay time
\( \tau  \). Depending on whether the input is subthreshold or suprathreshold,
the incoming signal induces a phase jump \begin{equation}
\label{eq:psp}
\phi _{j}((t+\tau )^{+}):=\min \{U^{-1}(U(\phi _{j}(t+\tau ))+\eps _{ji}),1\}
\end{equation}
which depends on the instantaneous phase \( \phi _{j}(t+\tau ) \)
of the postsynaptic oscillator and the coupling strength \( \eps _{ji} \).
The phase dependence is determined by a twice continuously differentiable
'potential' function \( U(\phi ) \) that is assumed to be strictly
increasing, \( U'(\phi )>0 \), concave (down), \( U''(\phi )<0 \),
and normalized such that \( U(0)=0 \), \( U(1)=1 \) (cf.\ \cite{Mirollo_et_al}). 

By choosing an appropriate function \( U \), this model is equivalent
(cf.\ \cite{TimmeChaos}) to different well known models of interacting
threshold elements. For instance, for the leaky integrate-and-fire
oscillator defined by the linear differential equation \( {}\dot{V}=I-V \)
\cite{Maass_et_al} (and threshold at \( V=1 \)), one obtains \( U_{\mathrm{IF}}(\phi )=I(1-e^{-T_{\mathrm{IF}}\phi }) \)
where \( T_{\mathrm{IF}}=\log (I/(I-1)) \) is the period of a non-interacting
oscillator and \( I>1 \) is a suprathreshold external current. Oscillators
described by nonlinear differential equations are covered by the Mirollo-Strogatz
approach, too. For instance, the conductance based threshold model
of a neuron \cite{Maass_et_al} leads to a different, more complicated
function \( U_{\mathrm{CB}}(\phi ) \) (for details see \cite{TimmeChaos}).
All analytical results presented here are derived for the above, general
class of interaction functions. In numerical investigations, we use
the functional form \( U_{\mathrm{IF}} \) but find qualitatively
similar results for different \( U \). In this Letter, we focus on
inhibitorily coupled networks (all \( \eps _{ij}\leq 0 \) and \( \eps _{ii}=0 \)). 

We perform a stability analysis of the synchronous state (\( \phi _{i}(t)=\phi _{0}(t) \)
for all \( i \)) that exists if the coupling strengths are normalized
such that \( \sum _{j\in \Pre (i)}\eps _{ij}=\eps <0 \). Its period
is given by \begin{equation}
\label{eq:period}
T=\tau +1-\beta _{0}
\end{equation}
 where \( \beta _{0}=U^{-1}(U(\tau )+\eps ) \). To construct a stroboscopic
map, a perturbation \( \boldsymbol {\delta }(0)\equiv \boldsymbol {\delta }=(\delta _{1},\ldots ,\delta _{N}) \)
of the phases, defined by \begin{equation}
\label{eq:define_delta}
\phi _{i}(0)=\phi _{0}(0)+\delta _{i}\, ,
\end{equation}
 is ordered according to the rank order \( \rank (\boldsymbol {\delta }) \)
of the \( \delta _{i} \): For each oscillator \( i \) we label the
perturbations \( \delta _{j} \) of its presynaptic oscillators \( j \)
(for which \( \eps _{ij}\neq 0 \)) according to their size\begin{equation}
\label{eq:spikeorder}
\Delta _{i,1}\geq \Delta _{i,2}\geq \ldots \geq \Delta _{i,k_{i}}
\end{equation}
where \noun{\( k_{i}:=|\Pre (i)| \)} is the number of its presynaptic
oscillators, called in-degree in graph theory \cite{Chartrand}. In
addition, we define \( \Delta _{i,0}=\delta _{i} \). For illustration
assume that an oscillator \( i \) has exactly two presynaptic oscillators
\( j_{1} \) and \( j_{2} \) such that \( \Pre (i)=\{j_{1},j_{2}\} \)
and \( k_{i}=2. \) For certain perturbations, oscillator \( i \)
first receives a signal from oscillator \( j_{2} \) and slightly
later from oscillator \( j_{1} \). This determines the rank order
(\( \delta _{j_{2}}>\delta _{j_{1}} \)) such that \( \Delta _{i,1}=\delta _{j_{2}} \)
and \( \Delta _{i,2}=\delta _{j_{1}} \). 

Using the phase shift function \( h(\phi ,\eps ):=U^{-1}(U(\phi )+\eps ) \)
and denoting \( D_{i,n}:=\Delta _{i,n-1}-\Delta _{i,n} \) for \( n\in \{1,\ldots ,k_{i}\} \)
we compute the time evolution of phase-perturbations \( \delta _{i}\ll 1 \),
starting near \( \phi _{0}(0)=\tau /2 \) without loss of generality.
The stroboscopic time-\( T \) map of the perturbations, \( \delta _{i}\mapsto \delta _{i}(T) \),
is obtained from the scheme 

\begin{tabular}{|c|c|}
\selectlanguage{english}
\( t \)
\selectlanguage{american}&
\selectlanguage{english}
\( \phi _{i}(t) \)
\selectlanguage{american}\\
\hline 
\selectlanguage{english}
\( \begin{array}{c}
0\\
\tauh -\Delta _{i,1}\\
\tauh -\Delta _{i,2}\\
\vdots \\
\tauh -\Delta _{i,k_{i}}\\
\tauh -\Delta _{i,k_{i}}\\
\qquad +1-\beta _{i,k_{i}}
\end{array} \)
\selectlanguage{american}&
\selectlanguage{english}
\( \begin{array}{c}
\frac{\tau }{2}+\delta _{i}=:\frac{\tau }{2}+\Delta _{i,0}\\
\hD {\tau }{1}\\
\hD {\beta _{i,1}}{2}\\
\vdots \\
\hD {\beta _{i,k_{i}-1}}{k_{i}}\\
\mbox {reset:}\, \, 1\mapsto 0\left. \begin{array}{c}
\, \\
\, 
\end{array}\right. 
\end{array} \)
\selectlanguage{american}\\
\end{tabular} 

\noindent where the right column gives the phases \( \phi _{i}(t) \)
of oscillator \( i \) at times \( t \) of pulse receptions or reset
given in the left column. Here the presynaptic oscillator from which
oscillator \( i \) receives the \( n^{\mathrm{th}} \) pulse during
this cycle is labeled by \( j_{n} \). The time to threshold \begin{equation}
\label{eq:time_to_spike}
T_{i}^{(0)}:=\frac{\tau }{2}-\Delta _{i,k_{i}}+1-\beta _{i,k_{i}}
\end{equation}
 is always smaller than the period \( T \). Hence the period-\( T \)
map of the perturbation can be expressed as \begin{equation}
\label{eq:delta_i}
\delta _{i}(T)=T-T_{i}^{(0)}-\frac{\tau }{2}=\beta _{i,k_{i}}-\beta _{0}+\Delta _{i,k_{i}}.
\end{equation}
Expanding \( \beta _{i,k_{i}} \) for small \( D_{i,n} \) one can
prove by induction that to first order \begin{equation}
\label{eq:beta_i}
\beta _{i,k_{i}}\doteq \beta _{0}+\sum _{n=1}^{k_{i}}p_{i,n-1}D_{i,n}
\end{equation}
 where \begin{equation}
\label{eq:fractions}
p_{i,n}:=\frac{U'(U^{-1}(U(\tau )+\sum _{m=1}^{n}\eps _{ij_{m}}))}{U'(U^{-1}(U(\tau )+\eps ))}
\end{equation}
 for \( n\in \{0,\, 1,\, \ldots ,k_{i}\} \). This results in a first
order map\begin{equation}
\label{eq:matrixequation}
\boldsymbol {\delta }(T)\doteq A\boldsymbol {\delta }
\end{equation}
 where the elements of the matrix \( A \) are given by \begin{equation}
\label{eq:matrixelements}
A_{ij}=\left\{ \begin{array}{ll}
p_{i,n}-p_{i,n-1} & \mbox {if}\, j=j_{n}\in \Pre (i)\\
p_{i,0} & \mbox {if}\, j=i\\
0 & \mbox {if}\, j\notin \Pre (i)\cup \{i\}.
\end{array}\right. 
\end{equation}
 Since \( j_{n} \) in (\ref{eq:fractions}) and (\ref{eq:matrixelements})
identifies the \( n^{\mathrm{th}} \) pulse received during this cycle
by oscillator \( i \), the first order operator depends on the rank
order of the perturbations, \( A=A(\rank (\boldsymbol {\delta })) \),
and the map \( A\boldsymbol {\delta } \) is piecewise linear. In
general, signals arriving almost simultaneously at the same oscillator
induce different phase changes, depending on the order of arrival:
For the above example of an oscillator \( i \) with exactly two presynaptic
oscillators \( j_{1} \) and \( j_{2} \) and equal coupling strengths,
\( \eps _{i,j_{1}}=\eps _{i,j_{2}}\,  \), the first of the two arriving
signals has a larger effect, encoded in the \( p_{i,n} \), than the
second, by virtue of the concavity of \( U(\phi ) \) (cf.\ Eq.\ (\ref{eq:psp})).
The respective matrix elements \( A_{i,j_{1}} \) and \( A_{i,j_{2}} \)
are differences between certain \( p_{i,n} \) and therefore have
different values depending on which signal is received first. This
is induced by the structure of the network together with the jump-like
interactions. For networks with homogeneous, global coupling different
matrices \( A \) can be identified by appropriately permuting the
oscillator indices. In general, however, this is impossible. 

Hence, in this stability problem, given a network structure, one generally
has to deal with an exponential number of operators instead of a single
stability matrix. We treat all these operators simultaneously: It
is straightforward to show that for all matrices \( A \) (independent
of the rank order of a perturbation and the parameters) the matrix
elements are non-negative, \( A_{ij}\geq 0 \). Due to time-translation
invariance all \( A \) are normalized row-wise, \( \sum _{j}A_{ij}=1 \)
for all \( i \), and exhibit a trivial eigenvalue \( \lambda _{1}=1 \).
Moreover, the diagonal elements are identical and smaller than one,
\( A_{ii}=A_{0}<1 \). The synchronous state is thus stable, because
the inequality

\begin{equation}
\label{eq:max_delta}
\left. \begin{array}{lll}
\max _{i}|\delta _{i}(T)| & \leq  & \max _{i}\sum _{j}|A_{ij}||\delta _{j}|\\
 & \leq  & \max _{i}\sum _{j}|A_{ij}|\max _{k}|\delta _{k}|=\max _{k}|\delta _{k}|,
\end{array}\right. 
\end{equation}
 is satisfied for all matrices \( A \). 

For a convex potential function \( U, \) where \( U''>0 \), and
excitatory interactions (\( \eps _{ij}\geq 0 \)) the synchronous
state is stable as well. The above proof applies if the total input
\( \eps  \) is not suprathreshold, i.e.\ \( \eps <1-U(\tau ) \).
Thus whereas excitatory interactions must not be too strong for applicability
of the proof, inhibititory interactions may be arbitrarily strong.

For structural stability of the stable synchronous state it is required
that the non-trivial eigenvalues of the matrices \( A \) are separated
from the unit circle. An instructive example is given by a network
of integrate-and-fire oscillators, \( U(\phi )=U_{\textrm{IF}}(\phi ) \),
where all matrices (\ref{eq:matrixelements}) are degenerate if \( \eps _{ij}=\eps /k_{i} \)
for all \( j\in \Pre (i) \). In this case, the eigenvalues of a single
matrix completely characterize the dynamics in the vicinity of the
synchronous state. Numerically, we find that in large random networks
in which every connection is present with probability \( p \) all
non-trivial eigenvalues are located in a disk \( D=\{z\in \mathbb {C}|\, |z-A_{0}|\leq r\} \)
of radius \( r \) that is centered at \( A_{0}<1 \) and separated
from the unit circle. An estimate for the radius, \( r=(1-A_{0})(p^{-1}-1)^{1/2}N^{-1/2} \)
for \( N\gg 1 \), can be obtained from the theory of \textit{Gaussian}
asymmetric \textit{}random matrices \cite{Sommers}. We find that
this estimate well agrees with our numerical results \cite{error}.
This indicates that in the limit of large \( N \), all non-trivial
eigenvalues are concentrated near \( z=A_{0} \) and thus separated
from unit circle.

\begin{figure}
{\centering \resizebox*{8cm}{!}{\includegraphics{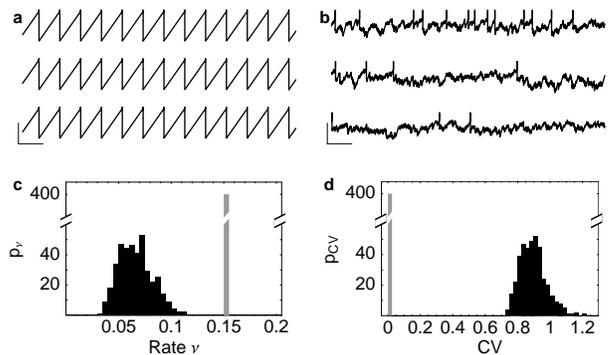}} \par}

\caption{Coexistence of (a) synchronous and (b) irregular dynamics in a random
network (\protect\( N=400\protect \), \protect\( p=0.2\protect \),
\protect\( I=4.0,\protect \) \protect\( \eps =16.0\protect \), \protect\( \tau =0.035\protect \)).
(a),(b): Trajectories of the potential \protect\( U(\phi _{i})\protect \)
of three oscillators (angular bars: time scale (horizontal) \protect\( \Delta t=8\protect \);
potential scales (vertical) (a) \protect\( \Delta U=8\protect \),
(b) \protect\( \Delta U=2\protect \) ; spikes of height \protect\( \Delta U=1\protect \)
added at firing times). (c),(d): Distributions (c) \protect\( p_{\nu }\protect \)
of rates and (d) \protect\( p_{\mathrm{CV}}\protect \) of the coefficient
of variation, displayed for the irregular (dark gray) and synchronous
(light gray) dynamics. \label{fig:trajectoryISIcv}}
\end{figure}

The above analysis shows that for inhibitory coupling the synchronous
state is stable, independent of the parameters and the network structure.
Numerical simulations show that for a network at given parameters
this synchronous state often coexists with one or more other attractors.
A particularly important example which occurs in randomly connected
networks with strong interactions, is a balanced state (cf.\ \cite{Brunel,VreeswijkScience})
that exhibits irregular dynamics. In this balanced state, found originally
in binary neural networks \cite{VreeswijkScience}, inhibitory and
excitatory inputs cancel each other on average but fluctuations lead
to a variability of the membrane potential and a high irregularity
in firing times (see also \cite{Brunel}). Figures \ref{fig:trajectoryISIcv}a,b
display sample trajectories of the potentials \( U(\phi _{i}) \)
of three oscillators for the same random network, making obvious the
two distinct kinds of coexisting dynamics.

The dynamical differences are quantified by a histogram \( p_{\nu } \)
of oscillator rates (Fig.\ \ref{fig:trajectoryISIcv}c) \begin{equation}
\label{eq:ISIdefinition}
\nu _{i}=\left( \left\langle t_{i,n+1}-t_{i,n}\right\rangle _{n}\right) ^{-1},
\end{equation}
 the reciprocal values of the time averaged inter-spike-intervals.
Here the \( t_{i,n} \) are the times when oscillator \( i \) fires
the \( n^{\mathrm{th}} \) time. The temporal irregularity of the
firing-sequence of single oscillators \( i \) is measured by the
coefficient of variation \begin{equation}
\label{eq:CVdefinition}
CV_{i}=\left( \nu _{i}^{2}\left\langle (t_{i,n+1}-t_{i,n})^{2}\right\rangle _{n}-1\right) ^{\frac{1}{2}},
\end{equation}
 defined as the ratio of the standard deviation of the inter-spike
intervals and their average. A histogram \( p_{\mathrm{CV}} \) of
the \( CV_{i} \) (Fig.\ \ref{fig:trajectoryISIcv}d) shows that the
irregular state exhibits coefficients of variation near one, the coefficient
of variation of a Poisson process. Such irregular states occur robustly
when changing parameters and network topology; on the other hand,
the size of the basin of attraction of the synchronous state is also
significant and increases with increasing delay \( \tau  \) . 

\begin{figure}
{\centering \resizebox*{7.5cm}{!}{\includegraphics{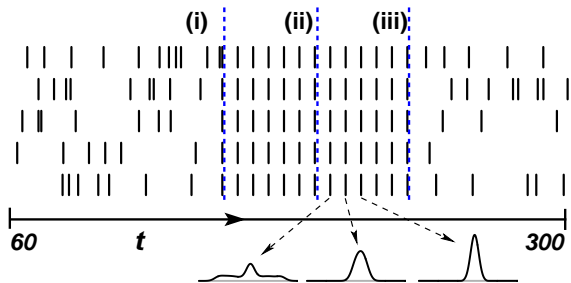}} \par}

\caption{Switching between synchronous and irregular dynamics (\protect\( N=400\protect \),
\protect\( p=0.2\protect \), \protect\( I=4.0\protect \), \protect\( \eps =16.0\protect \),
\protect\( \tau =0.14\protect \)). Firing times of five oscillators
are shown in a time window \protect\( \Delta t=240\protect \). Vertical
dashed lines mark external perturbations: (i) large excitatory pulses
lead to synchronous state, (ii) a small random perturbation (\protect\( |\Delta \phi _{i}|\leq 0.18\protect \))
is restored (iii) a sufficiently large random perturbation (\protect\( |\Delta \phi _{i}|\leq 0.36\protect \))
leads to an irregular state. Bottom: Time evolution of the spread
of the spike times after perturbation (ii), total length \protect\( \Delta t=0.25\protect \)
each. \label{fig:spike_perturb} }
\end{figure}

The coexistence of two qualitatively different kinds of dynamics leads
to the question how regular dynamics can be induced when the system
currently is in an irregular state and vice versa. A simple mechanism
to synchronize oscillators that are in a state of irregular firing
is the delivery of two sufficiently strong external excitatory (phase-advancing)
\textit{pulses} that are separated by a time \( \Delta t\in (\tau ,1) \),
cf.\ Fig.\ \ref{fig:spike_perturb}. The first pulse then leads to
a synchronization of phases due to simultaneous suprathreshold input
(cf.\ Eq.\ (\ref{eq:psp})) . If there are traveling signals that
have been sent but not received at the time of the first pulse, a
second pulse after a time \( \Delta t>\tau  \) is needed that synchronizes
the phases after all internal signals have been received. This synchronous
state is not affected by \textit{small random perturbations}, whereas
\textit{large random perturbations} lead back to irregular dynamics
(Fig.\ \ref{fig:spike_perturb}). Mechanisms for both directions of
switching may be realized in biological neural networks by external
neuronal populations: While strong external pulses may be generated
by external neurons that are highly synchronized, a random perturbation
can be realized by neurons which fire irregularly.

Most previous studies of the dynamics of networks of pulse-coupled
units focussed on regular networks or worked in some mean field limit
\cite{Abbott_et_al,Brunel,Dror,Mirollo_et_al}. These studies often
relied on the analysis of bifurcations from one state to another as
an external parameter is changed. Based on the stability analysis
developed here, that applies to networks with general connectivity,
we have demonstrated that regular synchronous dynamics may coexist
with irregular dynamics in sufficiently complex networks. The coexistence
of qualitatively different states at identical parameters indicates
that bifurcation approaches may often not give a complete picture
of the network dynamics, if the network structure is too complex.
This fact may well apply not only to networks of pulse-coupled units
but also to the dynamics of many other complex networks. In addition,
our results emphasize that in complex networks of pulse-coupled units
the occurrence of temporally regular and irregular firing patterns
may typically reflect the collective state of the network rather than
the dynamics of individual units.

The analysis presented in this Letter demonstrates that the dynamics
in certain complex networks can be revealed by considering the vertices
as units with simple dynamical properties, e.g.\ intrinsic oscillators.
Such systems provide promising starting points for future studies
addressing the \textit{dynamics} in networks, now that important aspects
of their \textit{complex structure} have been understood \cite{Albert_et_al}. 

We thank M. Diesmann, D. Hansel, M. Holicki, H. Sompolinsky, and C.
van Vreeswijk for helpful comments.

\end{document}